# Electronic properties of hexagonal tungsten monocarbide WC with 3*d* impurities from first-principles calculations

D.V. Suetin, I.R. Shein, A.L. Ivanovskii *

*Institute of Solid State Chemistry, Ural Branch of the Russian Academy of Sciences, Ekaterinburg GSP-145, 620041 Russia*

**Abstract**

First principles FLAPW- GGA calculations have been performed to predict the structural, electronic, cohesive and magnetic properties for hexagonal WC doped with all 3*d* metals. The optimized lattice parameters, density of states, cohesive and formation energies have been obtained and analyzed for ternary solid solutions with nominal compositions $W_{0.875}M_{0.125}C$ (where M = Sc, Ti….Ni, Cu). In addition, the magnetic properties of these solid solutions have been examined, and magnetization has been established for $W_{0.875}Co_{0.125}C$.

*PACS:* 61.72.Bb; 61.72.vV; 71.55.Gs; 71.15.Mb

*Keywords:* tungsten carbide; 3*d* impurities; electronic, cohesive, structural, magnetic properties; FLAPW-GGA calculations

## 1. Introduction

Tungsten monocarbide WC attracts much attention owing to its unique physical and chemical properties such as extreme hardness, high melting point, chemical inertness, interesting catalytic behavior *etc*, and belongs to the most promising engineering materials with a wide range of industrial applications [1-4]. Simultaneously, a much attention is given to crystalline and nano-sized WC-based composites comprising other transition metals (WC/M, where M are *d* metals) and related materials suitable for many technological applications.

* Corresponding author. Tel.: +7 3745331; fax: +7 373 3744495
 E-mail address: ivanovskii@ihim.uran.ru (Ivanovskii A.L.)

Today, for WC-M systems a set of various materials has been prepared.

An important group of such materials includes the above-mentioned WC/M composites (termed also as metallo-carbides or cemented carbides), namely, heterogeneous systems which consist of grains of tungsten carbide glued with a binder metal to combine the hardness of the carbide with the toughness of the metal, see [5-7].

Another group involves the so-called intermediate metal-rich phases such as $Fe_3W_3C$, $Co_3W_3C$ or $Co_6W_6C$. These phases (which adopt individual crystal structures and properties) may arise due to mutual solution in the interface region between WC and transition metals (or their alloys or carbides) or may be prepared using special synthetic routes [8,9].

Finally, a set of restricted solid solutions (SSs) $W_{1-x}M_xC$ (x << 1) may be formed, where *d* atoms are partially substituted for tungsten atoms. This situation can arise at contact of WC with metals or their carbides, during mechanical alloying from elemental powders or liquid-phase sintering, see [1-4,10,11].

While extensive theoretical studies have been performed for tungsten monocarbide WC [12-20], much less information is available for the electronic structure, stability and physical properties of the above-mentioned SSs $W_{1-x}M_xC$. To our knowledge, only WC with 25% substitution of Mo or Ti for tungsten and WC-12% M (M = Cr, Co, Zr) alloys have been considered in earlier works [10,11].

In this paper we have performed a systematic study of the effect of substitutional impurities - all transition 3*d* metal ions (Sc, Ti….Cu, Zn) on the structural, electronic, cohesive and magnetic properties of hexagonal WC.

**2. Models and computational method**

The basic phase in the W-C system is the hexagonal tungsten *mono*-carbide *h*-WC. This material has a hexagonal structure (space group *P*-6*m*2), where tungsten and carbon atoms form simple hexagonal layers, their stacking is of the *ABAB*…type. Both W and C sites are trigonal prismatic. On the other hand, systematic data on solid solutions in WC-M systems, where M are 3*d* transition metals are not available in the literature up to now.

In our simulations we examine the structural, electronic and magnetic properties of *h*-WC based SSs $W_{1-x}M_xC$ in the doping limit ~ 12%. For this purpose we use the periodic 16-atomic $W_8C_8$ supercell, in which one tungsten atom is replaced by M atoms (supercells $W_7MC_8$). In this way we model ternary SSs, which correspond to the nominal compositions of $W_{0.875}M_{0.125}C$.

Our band-structure calculations for all $W_{0.875}M_{0.125}C$ SSs (where M = Sc, Ti….Ni and Cu) were performed within the full potential method with mixed basis APW+lo (LAPW) implemented in the WIEN2k suite of programs [21]. The generalized gradient correction (GGA) to exchange-correlation potential of Perdew, Burke and Ernzerhof [22] was used. The electronic configurations were taken to be [Xe] $6s^25d^4$ for W, [Ar] $4s^23d^n$ for M and [He] $2s^22p^2$ for carbon. Here, the noble gas cores were distinguished from the sub-shells of valence electrons. The basis set inside each MT sphere was split into core and valence subsets. The core states were treated within the spherical part of the potential only, and were assumed to have a spherically symmetric charge density in muffin tin (MT) spheres. The valence part was treated with the potential expanded into spherical harmonics to $l = 4$. The valence wave functions inside the spheres were expanded to $l = 12$. The plane-wave expansion with $R_{MT} \times K_{MAX}$ was equal to 7, and *k* sampling with 10×10×10 *k*-points mesh in the Brillouin zone was used. Relativistic effects were taken into account within the scalar-relativistic approximation.

The self-consistent calculations were considered to have converged when the difference in the total energy of the crystal did not exceed 0.01 mRy as calculated at consecutive steps. In this way we have used a standard optimization regime as was described in original version WIEN2k [21]; this means a minimization of the total energy by variation on the lattice parameters (*a* and *c*) and the minimization of the atomic forces (< 0.05 mRy/a.u.). The density of states (DOS) was obtained using a modified tetrahedron method [23].

## 3. Results and discussion

### 3.1. Structural properties and density.

As the first step, total energy ($E_{tot}$) *versus* cell volume calculations were carried out for the hexagonal *mono*-carbide *h*-WC to determine the equilibrium structural parameters. The calculated values (*a* = 0.2926 nm and *c* = 0.2849 nm) are consistent with those measured experimentally (*a* = 0.2906 nm and *c* = 0.2837 nm, see [4]). Our theoretical values are slightly higher (at about 1%) as compared to the experimental data. This is a well known fact of overestimation of structural parameters for GGA calculations. The theoretical density (ρ, 15.395 g/cm$^3$) also agrees with experiment: ρ = 15.5-15.7 g/cm$^3$ [24].

Let us discuss the same parameters as obtained for the ternary $W_{0.875}M_{0.125}C$ SSs, Table 1 and Fig. 1. Note that our lattice parameters are in reasonable agreement with other available data. For example, our values for $W_{0.875}Ti_{0.125}C$ are *a* = 0.2927 nm and *c/a* = 0.9696 - as compared with *a* = 0.289 nm and *c/a* = 0.961 for $W_{0.750}Ti_{0.250}C$ [10]. According to our FLAPW-GGA calculations, the parameters *a* and *c/a* for $W_{0.875}Cr_{0.125}C$ and $W_{0.875}Co_{0.125}C$ are 0.2905 nm, 0.972 and 0.2912 nm, 0.968, respectively, as compared with *a* = 0.2905 nm and *c/a* =0.972 (for $W_{0.875}Cr_{0.125}C$) and *a* = 0.2901 nm and *c/a* = 0.976 (for $W_{0.875}Co_{0.125}C$) as obtained by means of the LDA approach (VASP code) [11]. What is more interesting, a non-monotonous behavior of *a* and *c* parameters was established in our systematic calculations as going from $W_{0.875}Sc_{0.125}C$ to $W_{0.875}Cu_{0.125}C$. In addition, for various $W_{0.875}M_{0.125}C$ SSs the trends of *a* or *c* parameter variation were different, Table 1. Really, the parameter *a* decreased from $W_{0.875}Sc_{0.125}C$ to $W_{0.875}Cr_{0.125}C$ and then increased as going to $W_{0.875}Cu_{0.125}C$; whereas the minimal value for the constant $c$ was obtained for $W_{0.875}Fe_{0.125}C$, see Table 1. Thus, *anisotropic deformation* of the crystal structure of the matrix (hexagonal WC) took place when tungsten was replaced by 3*d* atoms.

Naturally, it should be expected that all $W_{0.875}M_{0.125}C$ alloys containing more light 3*d* atoms will adopt lower density as compared with WC. Really, the lowering of ρ change from 6 % (for $W_{0.875}Fe_{0.125}C$) to 11% (for $W_{0.875}Sc_{0.125}C$). However if the atomic masses of substitutional M atoms increase as going from Sc to Cu, the cell volumes (V) change non-monotonously, and the density of the solid solutions also changes non-monotonously and adopts the maximal values for compositions with Mn and Fe, see Fig. 1.

*3.2. Cohesive properties and energies of formation.*

To provide an insight in the fundamental aspects of phase relations in the W-M-C systems, some estimations based on the total energy calculations were performed.

For this purpose, we have calculated the cohesive energy ($E_{coh}$) for $W_{0.875}M_{0.125}C$ SSs as $E_{coh} = E_{tot}(W_{0.875}M_{0.125}C) - \{0.875\ E_{tot}(W^{at}) + 0.125\ E_{tot}(M^{at}) + E_{tot}(C^{at})\}$, where $E_{tot}(W_{0.875}M_{0.125}C)$, $E_{tot}(W^{at})$, $E_{tot}(M^{at})$ and $E_{tot}(C^{at})$ are the total energies of the $W_{0.875}M_{0.125}C$ SSs and free W, M and carbon atoms, respectively. Our results (see Fig. 2) show that $E_{coh}$ for all the solid solutions are smaller than that for the pure *mono*-carbide *h*-WC ($E_{tot}$(WC = -21.38 eV/cell)). This means that the introduction of all 3*d* metal impurities destabilizes the WC structure. Besides, the effect of such destabilization depends on type of the impurity and is maximal for metals of the end of the 3*d* row, which, as is known [1-3] from reactions with carbon, do not form stable *mono*-carbides.

The same tendency has been obtained in our calculations of the formation energies $E_{form}$. Here, to estimate the relative stability of the modeled $W_{0.875}M_{0.125}C$ SSs, their free Gibbs energies ($G = ΔH + P·V - T·S$) should be established. Since our calculations are performed at zero temperature and zero pressure, $G$ becomes equal to the enthalpy of formation $\Delta H$, which can be estimated from the total energies of $W_{0.875}M_{0.125}C$ SSs and their constituents. The reactions of phase formation from simple substances are usually used for the estimation of $\Delta H$. Thus, the formation energies of $W_{0.875}M_{0.125}C$ with respect to the *bcc* tungsten, 3*d* metals and graphite ($C^g$) in the formal reactions $W + M + C^g \rightarrow W_{0.875}M_{0.125}C$ are obtained in our FLAPW-GGA calculations as: $E_{form}(W_{0.875}M_{0.125}C) = E_{tot}(W_{0.875}M_{0.125}C) - \{0.875\ E_{tot}(W^{met}) + 0.125\ E_{tot}(M^{met}) + E_{tot}(C^g)\}$, where $E_{tot}(W^{met})$, $E_{tot}(M^{met})$ and $E_{tot}(C^g)$ are the total energies of W, 3*d* metals and graphite at their optimized geometries, respectively. Thus, a negative $E_{form}$ indicates that it is energetically favorable for given reagents to mix and form stable ternary carbide phases, and *vice versa*.

The results shown in Fig. 2 bring us to the following conclusions. For the *mono*-carbide, the maximal negative $E_{form}(h\text{-WC}) = -0.34$ eV/f.u. was obtained. Thus, these data indicate that the binary *h*-WC has the highest stability among all the considered species. For $W_{0.875}M_{0.125}C$ SSs with the metals of the beginning of the 3*d* row, their $E_{form}$ are negative, whereas the formation energies of $W_{0.875}M_{0.125}C$ SSs with the metals of the end of the 3*d* row are positive, and therefore these systems should be unstable. For $W_{0.875}(Fe,Co,Ni)_{0.125}C$ SSs, their formation energies are close to zero, and these systems should be treated as meta-stable. Naturally, thermodynamic factors such as temperature and pressure should be taken into account while comparing these data with experiments.

*3.3. Electronic structure and magnetic properties.*

To understand the electronic properties of $W_{0.875}M_{0.125}C$ SSs, let us briefly discuss the band structure of the matrix phase – the binary *h*-WC, see also [12-20]. The density of states (DOS) of the *h*-WC shown in Fig. 3 is in good agreement with earlier data [12-20]. The valence DOS can be divided into three main regions (peaks A-C, Fig. 3); (i) from -14.5 eV to -10 eV with mainly carbon 2*s* states, (ii) from − 7.8 eV to -2.5 eV with strongly hybridized W 5*d* and C *2s* states, and finally (iii) from -2.5 eV to $E_F$ = 0 eV with mainly tungsten 5*d* states. The Fermi level $E_F$ is close to the DOS minimum, which qualitatively points to high stability of this material. The bonding picture for the *h*-WC as derived from these data can be classified as a mixture of metallic, ionic, and covalent contributions. The metallic contribution can be attributed to the partially filled W 5*d* states, peak C Fig. 3; the presence of a strong covalent bonding is attested by hybridization of the C 2*p* and W 5*d* states (peaks B and C), and, finally, the ionic nature of the W-C bonds is determined by partial transfer of charge density from tungsten to carbon atoms due to differences in their elecronegativity values.

Let us discuss the electronic structure of $W_{0.875}M_{0.125}C$ SSs. Note that for all $W_{0.875}M_{0.125}C$ our calculations were performed in two variants: for nonmagnetic (NM) and magnetic states (in approximation of ferromagnetic ordering). For all ternary solid solutions except $W_{0.875}Co_{0.125}C$ the calculations of the magnetic states converged into NM states.

Figure 4 shows the total and partial DOSs for all NM $W_{0.875}M_{0.125}C$ SSs; the band structure parameters (band width) for these materials are presented in Table 2. The DOS picture shows that as going from Sc to Cu: (i) impurity M 3*d* states move in the VB region and these states begin to fill, and (ii) the filling of the common valence band of $W_{0.875}M_{0.125}C$ SSs increase owing to an increasing number of *d* electrons. As a result: (i) the common VB width increases from 13.88 eV ($W_{0.875}Sc_{0.125}C$) to 14.66 eV ($W_{0.875}Cu_{0.125}C$); (ii) the Fermi level moves from the bottom of the hybridized (W 5*d* - M 3*d* - C2*p*) band to the local DOS minimum - between bonding (*p-d*) and anti-bonding (W 5*d* + M3*d*) bands, and (iii) a quite non-monotonous change in the near-Fermi DOS ($N(E_F)$, see Table 3) as going from Sc to Cu is determined by the two mentioned above factors, namely, an increased band filling and movement of $E_F$. Note that the minimal value of $N(E_F)$ was obtained for $W_{0.875}Cr_{0.125}C$ (i.e. at isoelectronic substitution W → Cr), which is the most stable solid solution (according to our estimations of $E_{coh}$ and $E_{form}$).

The obtained values of N($E_F$) allow us to estimate the Sommerfeld constants ($\gamma$) and the Pauli paramagnetic susceptibility ($\chi$), assuming the free electron model, as: $\gamma = (\pi^2/3)N(E_F)k^2_B$, and $\chi = \mu_B^2 N(E_F)$, Table 3. The results obtained indicate that the concentration of the delocalized near-Fermi electrons of $W_{0.875}M_{0.125}C$ SSs can be smaller (for $W_{0.875}Cr_{0.125}C$, $W_{0.875}Ni_{0.125}C$ and $W_{0.875}Mn_{0.125}C$) or greater (for all other SSs) than for the starting binary phase h-WC depending on the 3d metal type.

Finally, the obtained spin-resolved DOS for the magnetic $W_{0.875}Co_{0.125}C$ is shown in Fig. 5. The magnetism takes its origin in the spin polarization of Co 3d states (the calculated local magnetic moment of cobalt is about 0.12 $\mu_B$), whereas the induced magnetization on the nearest tungsten and carbon atoms of the matrix is very small.

## 4. Conclusions

In conclusion, the *ab initio* band structure calculations have been performed to understand the structural, cohesive, electronic, and magnetic properties of a series of h-WC based solid solutions with the nominal composition $W_{0.875}M_{0.125}C$ as a function of the 3d metal type (M = Sc, Ti….Ni and Cu).

The results obtained are summarized as follows:

(i) a non-monotonous anisotropic deformation of the crystal structure of the matrix (h-WC) takes place when tungsten is replaced by various 3d atoms; the density of $W_{0.875}M_{0.125}C$ solid solutions also changes non-monotonously and adopts the maximal values for $W_{0.875}Mn_{0.125}C$ and $W_{0.875}Fe_{0.125}C$;

(ii) the cohesive ($E_{coh}$) and formation energies ($E_{form}$) of $W_{0.875}M_{0.125}C$ SSs were estimated. The findings obtained show that all the SSs are less stable than the binary *mono*-carbide h-WC. At the same time, the most stable SS among them should be $W_{0.875}Cr_{0.125}C$ with isoelectronic substitution W → Cr.

(iii) the electronic structure of $W_{0.875}M_{0.125}C$ SSs is determined mainly by the band filling variation and movement of $E_F$. Note that in the series of the examined SSs the DOS values at the Fermi level N($E_F$) (as well as the values of such parameters as the Sommerfeld constants ($\gamma$) and the Pauli paramagnetic susceptibility ($\chi$)) can be greater or smaller than the corresponding values for the *h*-WC depending on the type of the 3*d* metal.

(vi) all the considered $W_{0.875}M_{0.125}C$ SSC are non-magnetic, except $W_{0.875}Co_{0.125}C$, for which the magnetic state is more energetically favorable. Our results show that the magnetism for $W_{0.875}Co_{0.125}C$ is due to spin polarization of Co 3*d* states, and the local magnetic moment of cobalt is about 0.12 $\mu_B$, whereas the induced magnetization on the nearest tungsten and carbon atoms of the matrix is very small.

**Acknowledgements.** The work was supported by RFBR, grant No. 08-08-00034-a.

**TABLES**

Table 1.
Optimized lattice parameters (*a* and *c*, in nm) for ternary $W_{0.875}M_{0.125}C$ alloys in comparison with hexagonal *mono*-carbide WC.

| phase | WC | $W_{0.875}Sc_{0.125}C$ | $W_{0.875}Ti_{0.125}C$ | $W_{0.875}V_{0.125}C$ | $W_{0.875}Cr_{0.125}C$ |
|---|---|---|---|---|---|
| *a* | 0.2926 | 0.2952 | 0.2927 | 0.2912 | 0.2905 |
| *c* | 0.2849 | 0.2856 | 0.2838 | 0.2830 | 0.2824 |
| phase | $W_{0.875}Mn_{0.125}C$ | $W_{0.875}Fe_{0.125}C$ | $W_{0.875}Co_{0.125}C$ | $W_{0.875}Ni_{0.125}C$ | $W_{0.875}Cu_{0.125}C$ |
| *a* | 0.2905 | 0.2909 | 0.2912 | 0.2920 | 0.2929 |
| *c* | 0.2818 | 0.2812 | 0.2819 | 0.2820 | 0.2828 |

Table 2.
Calculated band structure parameters (band width, in eV) for ternary $W_{0.875}M_{0.125}C$ alloys in comparison with hexagonal *mono*-carbide WC.

| | Band types | | | |
|---|---|---|---|---|
| phase | Common band (C 2*s* ÷ $E_F$) | C 2*s* band | gap C 2*s* ÷ C 2*p* + W 5*d* | Valence band C 2*p* + W 5*d* (to $E_F$) |
| $W_{0.875}Sc_{0.125}C$ | 13.88 | 4.07 | 1.85 | 7.96 |
| $W_{0.875}Ti_{0.125}C$ | 13.97 | 4.13 | 1.74 | 8.10 |
| $W_{0.875}V_{0.125}C$ | 14.08 | 4.13 | 1.72 | 8.23 |
| $W_{0.875}Cr_{0.125}C$ | 14.38 | 4.14 | 1.72 | 8.52 |
| $W_{0.875}Mn_{0.125}C$ | 14.73 | 4.15 | 1.74 | 8.84 |
| $W_{0.875}Fe_{0.125}C$ | 14.76 | 4.15 | 1.75 | 8.86 |
| $W_{0.875}Co_{0.125}C$* | 14.65 | 4.15 | 1.73(↑), 1.71(↓) | 8.77 (↑), 8.75 (↓) |
| $W_{0.875}Ni_{0.125}C$ | 14.59 | 4.18 | 1.70 | 8.71 |
| $W_{0.875}Cu_{0.125}C$ | 14.66 | 4.16 | 1.68 | 8.82 |
| WC | 14.51 | 4.22 | 1.95 | 8.34 |

* for magnetic $W_{0.875}Co_{0.125}C$ – for spin up (↑) and spin down (↓) bands.

Table 3.
Total $N(E_F)$ and partial $N^l(E_F)$ density of states at the Fermi level (in states/eV·cell·spin), electronic heat capacity γ (in mJ·K$^{-2}$·mol$^{-1}$) and molar Pauli paramagnetic susceptibility χ (in $10^{-4}$ emu/mol) for ternary non-magnetic $W_{0.875}M_{0.125}C$ alloys in comparison with hexagonal *mono*-carbide WC.

| phase | $N^{tot}(E_F)$ | $N^{W5d}(E_F)$ | $N^{C2p}(E_F)$ | $N^{M3d}(E_F)$ | γ | χ |
|---|---|---|---|---|---|---|
| $W_{0.875}Sc_{0.125}C$ | 0.351 | 0.107 | 0.054 | 0.016 | 0.83 | 0.11 |
| $W_{0.875}Ti_{0.125}C$ | 0.442 | 0.132 | 0.058 | 0.043 | 1.05 | 0.14 |
| $W_{0.875}V_{0.125}C$ | 0.239 | 0.063 | 0.033 | 0.030 | 0.57 | 0.07 |
| $W_{0.875}Cr_{0.125}C$ | 0.147 | 0.045 | 0.019 | 0.021 | 0.35 | 0.05 |
| $W_{0.875}Mn_{0.125}C$ | 0.215 | 0.088 | 0.010 | 0.048 | 0.51 | 0.07 |
| $W_{0.875}Fe_{0.125}C$ | 0.314 | 0.108 | 0.024 | 0.090 | 0.74 | 0.14 |
| $W_{0.875}Ni_{0.125}C$ | 0.181 | 0.055 | 0.035 | 0.029 | 0.43 | 0.06 |
| $W_{0.875}Cu_{0.125}C$ | 0.270 | 0.085 | 0.051 | 0.025 | 0.64 | 0.08 |
| WC | 0.228 | 0.099 | 0.029 | - | 0.54 | 0.07 |

**FIGURES**

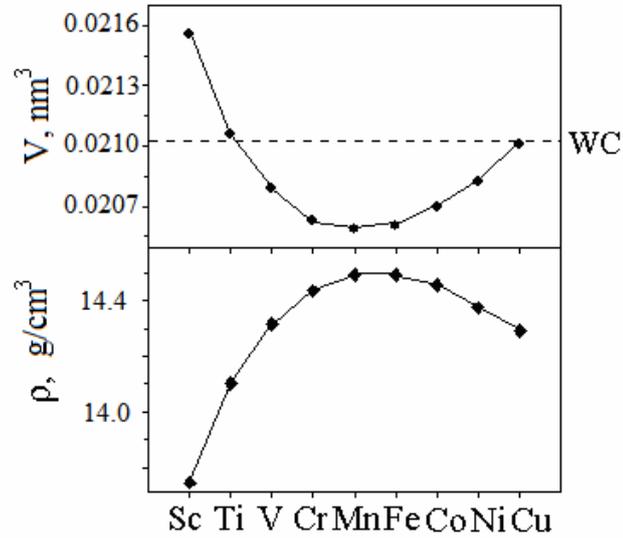

Fig. 1. Cell volumes (V, nm$^3$) and density (ρ, g/cm$^3$) for ternary $W_{0.875}M_{0.125}C$ SSs. Cell volume for hexagonal *mono*-carbide *h*-WC is also given.

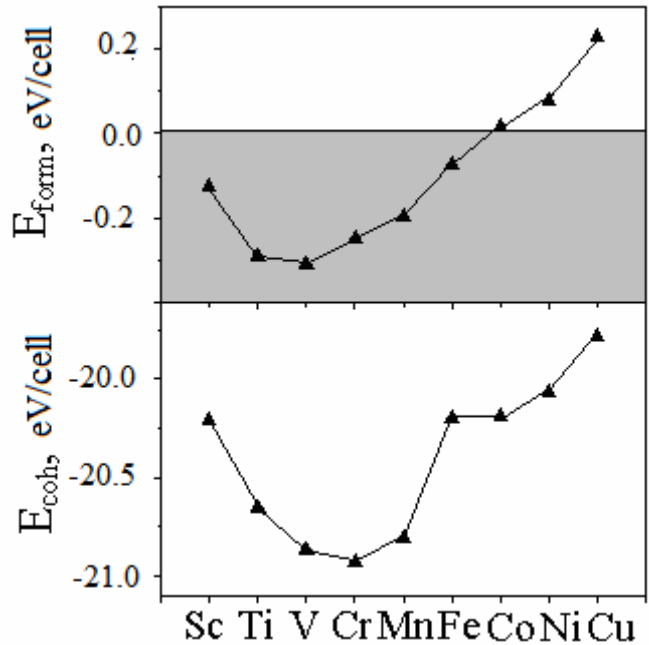

Fig. 2. Formation ($E_{form}$) and cohesive energies ($E_{coh}$) for ternary $W_{0.875}M_{0.125}C$ SSs.

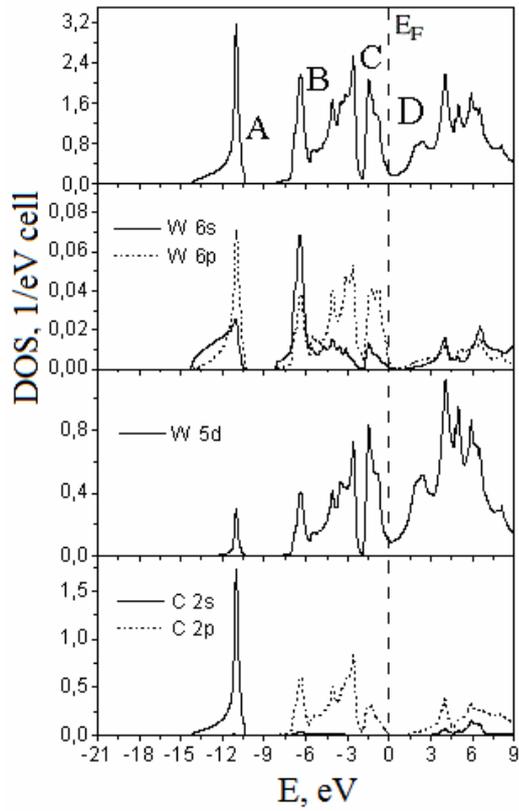

Fig. 3. The total (*upper panel*) and partial density of states for *h*-WC.

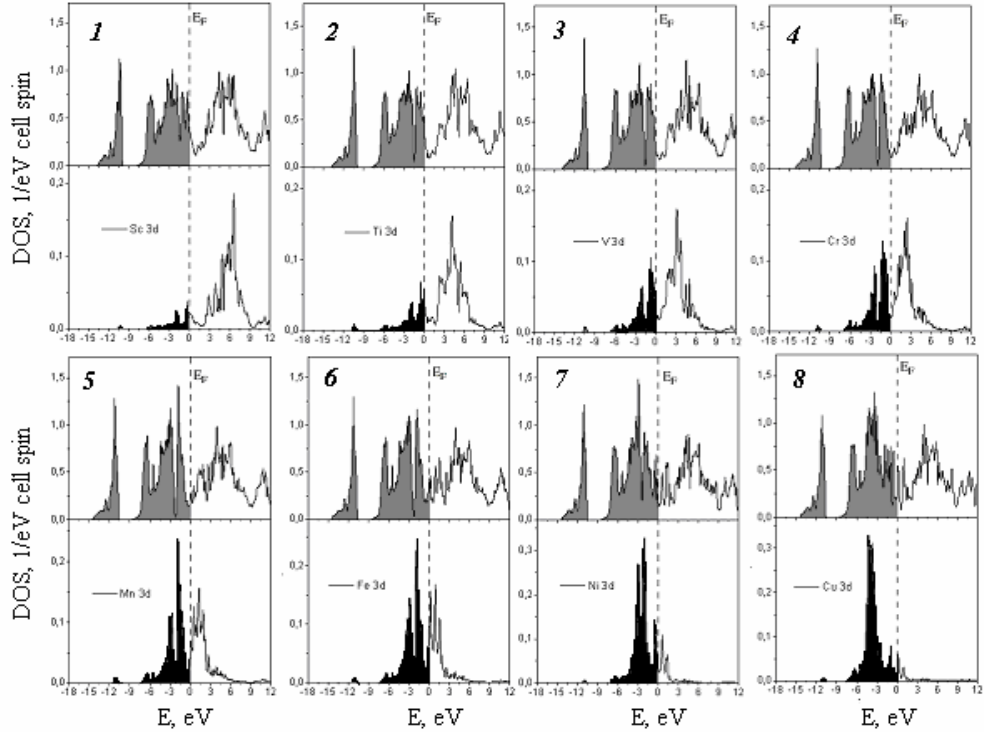

Fig. 4. The total (*upper panels*) and partial M 3*d* density of states for ternary non-magnetic $W_{0.875}M_{0.125}C$ SSs. 1- $W_{0.875}Sc_{0.125}C$; 2- $W_{0.875}Ti_{0.125}C$; 3- $W_{0.875}V_{0.125}C$; 4- $W_{0.875}Cr_{0.125}C$; 5- $W_{0.875}Mn_{0.125}C$; 6- $W_{0.875}Fe_{0.125}C$; 7- $W_{0.875}Co_{0.125}C$; 8- $W_{0.875}Ni_{0.125}C$.

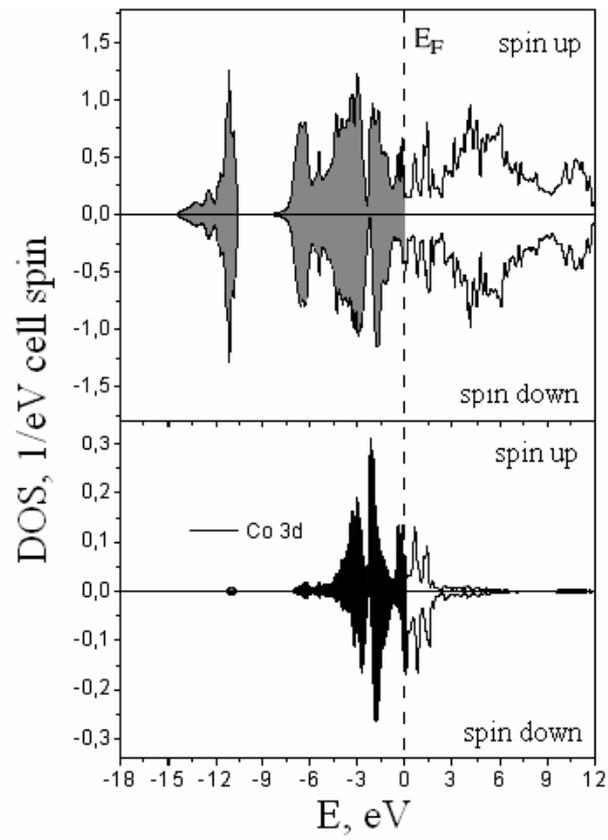

Fig. 5. The spin-resolved total (*upper panel*) and partial Co 3*d* density of states for ternary magnetic $W_{0.875}Co_{0.125}C$ SS.